\documentclass[prd,preprint,superscriptaddress,showpacs,byrevtex]{revtex4}
\usepackage{bm}
\def\fsl#1{\setbox0=\hbox{$#1$}           
   \dimen0=\wd0                                 
   \setbox1=\hbox{/} \dimen1=\wd1               
   \ifdim\dimen0>\dimen1                        
      \rlap{\hbox to \dimen0{\hfil/\hfil}}      
      #1                                        
   \else                                        
      \rlap{\hbox to \dimen1{\hfil$#1$\hfil}}   
      /                                         
   \fi}                                         %
\newcommand{\be}{\begin{equation}}
\newcommand{\ee}{\end{equation}}
\newcommand{\bea}{\begin{eqnarray}}
\newcommand{\eea}{\end{eqnarray}}
\newcommand{\beq}{\begin{equation}}
\newcommand{\eeq}{\end{equation}}
\newcommand{\beqs}{\begin{eqnarray}}
\newcommand{\eeqs}{\end{eqnarray}}
\newcommand{\lsim}{\mathrel{\raisebox{-
.6ex}{$\stackrel{\textstyle<}{\sim}$}}}

\newcommand{\aslash}{A\hspace{-0.067in}\slash}

\begin{document}
\title{ Matter-Antimatter Asymmetry Of The Universe and Baryon Formation From Non-Equilibrium Quarks and Gluons }
\author{Gouranga C Nayak }\thanks{E-Mail: nayakg138@gmail.com}
%
%
\date{\today}
\begin{abstract}
Baryon number violation, CP violation and non-equilibrium evolution of the early universe just after the big bang are proposed in the literature to be necessary conditions to explain the observed matter-antimatter asymmetry (baryon asymmetry) of the universe. Since the free quarks are not observed and the baryons are formed by the confinement of quarks and gluons, it is necessary to understand how the baryons were formed for the first time from the non-equilibrium quarks and gluons in the early universe just after the big bang in order to study the baryon asymmetry of the universe. In this paper we study the baryon formation from the non-equilibrium quarks and gluons in the early universe just after the big bang from the first principle in QCD by using the closed-time path integral formalism in non-equilibrium QCD.
\end{abstract}
\maketitle
\pagestyle{plain}

\pagenumbering{arabic}

\section{ Introduction }

Our universe began with the big bang. Although the laws of physics before the Planck time ($\sim 10^{-43}$ seconds of the big bang) is not known but
most of the known laws of physics are applicable after the Planck time. The temperature of the universe near the Planck time was $\sim 10^{19}$ GeV ($\sim 10^{32}$K) where the length scale of the universe was $\sim 10^{-35}$ meter.

At the Planck length scale it is necessary to study the quantum theory of gravity. However, unlike quantum electrodynamics (QED), quantum chromodynamics (QCD) and quantum theory of weak interaction which are renormalizable, the quantum gravity is not renormalizable. Due to this reason our known laws of physics are not applicable before the Planck time of the universe.

The four fundamental forces of the nature are: 1) strong force (nuclear force or QCD force), 2) electromagnetic force, 3) weak force and 4) gravitational force. Before the Planck time of the universe all these four fundamental forces of the nature could have unified to a single force where the laws of the quantum physics could be the theory of everything (TOE). However, since the quantum gravity is not renormalizable, we cannot formulate a theory of everything (TOE) at present.

After the Planck time of the big bang the different phases of the early universe which are of interest to the particle physics aspects of the phase transitions are, 1) the grand unified theory (GUT) phase transition, 2) the electro-weak phase transition and 3) the quark-gluon plasma phase transition.

The GUT time scale corresponds $\sim 10^{-36}$ seconds of the big bang with the temperature $\sim 10^{16}$ GeV ($\sim 10^{28}$K). At the GUT energy scale the strong force, the electromagnetic force and the weak force were unified where the quantum chromodynamics (QCD) coupling, the quantum electrodynamics (QED) coupling and the weak coupling merged to a single coupling. After this time the GUT symmetry was broken which created the strong force (nuclear force or QCD force) and the electro-weak force.

The electro-weak time scale is $\lsim 10^{-12}$ seconds of the big bang with the temperature $\sim 245$ GeV ($\sim 10^{15}$K). After this time scale the electro-weak force was spontaneously broken by the Higgs mechanism giving the masses to quarks, leptons and gauge bosons. Two separate forces (the electromagnetic force and the weak force) originated from this spontaneously broken symmetry of the electro-weak force.

The quark-gluon plasma time scale is $\sim 10^{-12}- 10^{-4}$ seconds after the big bang with the temperature $\sim 200$ MeV ($\sim 10^{12}K$). The quark-gluon plasma consists of quarks and gluons which are the fundamental particles of the nature. The strong force (the nuclear force or the QCD force) is responsible for the interaction between quarks and gluons.

Hadronization of the quark-gluon plasma occurred after $\sim 10^{-4}$ seconds of the big bang when the temperature of the universe was $< 200$ MeV ($< 10^{12}K$). The nucleons (proton and neutron) of the universe were formed for the first time from the quark-gluon plasma at this stage of the evolution of the universe. The proton and neutron are known as baryons and the antiproton and antineutron are known as antibaryons.

From the astrophysical experiments it is found that our universe contains more matter (baryons) than the antimatter (antibaryons). If $n_b$ is the number density of the baryons, $n_{\bar b}$ is the number density of the antibaryons and $n_\gamma$ is the number density of the photons then the baryon asymmetry parameter $\eta$ is defined by
\bea
\eta=\frac{n_b-n_{\bar b}}{n_\gamma}=\frac{n_B}{n_\gamma}
\label{eta}
\eea
where $n_B=n_b-n_{\bar b}$ is the net baryon number density.
The WMAP experiment \cite{bs} found that the net baryon to photon ratio of our universe is
\bea
\eta \sim 6.1 \times 10^{-10}.
\label{bas}
\eea
This experimental evidence of more matter (baryons) than antimatter (antibaryons) in our universe is known as the matter-antimatter asymmetry (or the baryon asymmetry) of the universe.

In order to explain the matter-antimatter asymmetry (the baryon asymmetry) of the universe it is proposed that the following three conditions
were satisfied during the early universe just after the big bang \cite{bs2}. These three conditions are: 1) baryon number violation at the GUT energy scale, 2) CP violation at the electro-weak energy scale and 3) non-equilibrium evolution of the early universe just after the big bang.

The baryons (such as proton and neutron) and antibaryons (such as antiproton and antineutron) are different from quarks and gluons because quarks and gluons carry color charges but the baryons/antibaryons do not carry color charges, {\it i. e.}, baryons/antibaryons are colorless. Since we have not directly experimentally observed quarks and gluons what we experimentally observe are the hadrons (such as baryons/antibaryons). The quarks and gluons are confined inside the hadron (such as inside the baryon/antibaryon) due to confinement in QCD [a phenomenon which is absent in QED]. Hence it is necessary to understand how the baryons/antibaryons were formed from the non-equilibrium quarks and gluons during the early universe just after $\sim 10^{-4}$ seconds of the big bang in order to study the baryon asymmetry of the universe.

As mentioned above the QCD \cite{ymr} is the fundamental theory of the nature to study the interaction between the quarks and gluons (the partons). The asymptotic freedom in QCD \cite{gwr} allows us to calculate the short distance partonic scattering cross section by using the perturbative QCD (pQCD). Using the factorization theorem in QCD \cite{fcr,fcr1,fcr2} the hadronic cross section can be calculated at the high energy colliders by using the experimentally extracted parton distribution function (PDF) and fragmentation function (FF).

The formation of baryons (such as proton and neutron) and antibaryons (such as antiproton and antineutron) from the quarks and gluons cannot be studied by using the pQCD because the formation of baryon/antibaryon from the quarks and gluons is a low energy phenomenon in QCD where the QCD coupling becomes large. Hence the non-perturbative QCD is necessary to study the baryon/antibaryon formation from the quarks and gluons from the first principle.

As mentioned above, in order to explain the baryon asymmetry of the universe, it is necessary to understand how the baryons/antibaryons were formed for the first time from the non-equilibrium quarks and gluons during the early universe just after $\sim 10^{-4}$ seconds of the big bang. Since the baryon/antibaryon formation from the quarks and gluons can be studied from the first principle by using the non-perturbative QCD one finds that the baryon/antibaryon formation from the non-equilibrium quarks and gluons can be studied from the first principle by using the nonequilibrium-nonperturbative QCD. The non-equilibrium QCD can be studied from the first principle by using the closed-time path integral formalism in quantum field theory \cite{ne,ne1,ne2,ne3}. Hence we need to study the nonequilibrium-nonperturbative QCD from the first principle to understand the baryons/antibaryons formation from the non-equilibrium quarks and gluons during the early universe just after the big bang.

In this paper we study the baryons (proton and neutron) and the antibaryons (antiproton and antineutron) formation from the non-equilibrium quarks and gluons in the early universe just after the big bang from the first principle in QCD by using the closed-time path integral formalism in the non-equilibrium QCD. The extension of this formalism  to other baryons/antibaryons is straightforward.

The paper is organized as follows. In section II we describe the baryons (proton and neutron) and the antibaryons (antiproton and antineutron)  formation from the quarks and gluons in QCD in vacuum. In section III we describe the baryons (proton and neutron) and the antibaryons (antiproton and antineutron) formation from the finite temperature quarks and gluons in QCD. In section IV we study the baryons (proton and neutron) and the antibaryons (antiproton and antineutron) formation from the non-equilibrium quarks and gluons in the early universe just after the big bang from the first principle in QCD by using the closed-time path integral formalism in the non-equilibrium QCD. Section V contains conclusions.

\section{Baryon and Antibaryon formation from quarks and gluons in QCD in vacuum}

The generating functional in QCD in the absence of any external sources is given by
\bea
&& Z[0]=\int [d{\bar \psi}_U][d\psi_U] [d{\bar \psi}_D] [d\psi_D] \times {\rm det}[\frac{\delta G_F^d}{\delta \omega^c}] \times {\rm exp}[i\int d^4x [-\frac{1}{4} F_{\mu \lambda}^b(x)F^{\mu \lambda b}(x)-\frac{1}{2\alpha} [G_F^b(x)]^2 \nonumber \\
&& +{\bar \psi}^j_U(x)[\delta^{jl}(i{\not \partial} -m_U)+gT^b_{jl}\aslash^b(x)]\psi^l_U(x) +{\bar \psi}^j_D(x)[\delta^{jl}(i{\not \partial} -m_D)+gT^b_{jl}\aslash^b(x)]\psi^l_D(x)]]\nonumber \\
\label{ab}
\eea
where $\psi_U^k(x)$ is the quark field for the up ($U$) quark, the $\psi_D^k(x)$ is the quark field for the down ($D$) quark, $k=1,2,3$ is the color index of the quark field, $\alpha$ is the gauge fixing parameter, $G_F^b(x)$ is the gauge fixing term, $m_U$ ($m_D$) is the mass of the up (down) quark, $A_\lambda^b(x)$ is the gluon field with Lorentz index $\lambda=0,1,2,3$, the color index $b=1,...,8$ and
\bea
F_{\mu \lambda}^b(x) =\partial_\mu A_\lambda^b(x) -\partial_\lambda A_\mu^b(x) +gf^{bds} A_\mu^d(x) A_\lambda^s(x).
\label{bb}
\eea
We do not have ghost fields in the QCD generating functional in eq. (\ref{ab}) as we directly work with the ghost determinant ${\rm det}[\frac{\delta G_F^d}{\delta \omega^c}]$ in this paper.

The partonic operator ${\cal O}_B(x)$ for the formation of baryon/antibaryon $B$ (where $B=p,n,{\bar p},{\bar n}$= proton, neutron, antiproton, antineutron) is given by
\bea
&& {\cal O}_p(x) =\epsilon^{jlk} \psi^{jT}_C\gamma_5 \psi^l_D \psi^k_U(x),~~~~~~~~~~~~~~~~{\cal O}_n(x) =\epsilon^{jlk} \psi^{jT}_DC\gamma_5 \psi^l_U \psi^k_D(x), \nonumber \\
&& {\cal O}_{\bar p}(x) =\epsilon^{jlk} {{\bar \psi}^{jT}_U}C\gamma_5 {\bar \psi}^l_D {\bar \psi}^k_U(x),~~~~~~~~~~~~~~~~{\cal O}_{\bar n}(x) =\epsilon^{jlk} {{\bar \psi}^{jT}_D}\gamma_5 {\bar \psi}^l_U {\bar \psi}^k_D(x)
\label{cb}
\eea
where $C$ is the charge conjugation operator.

The vacuum expectation value of the two point non-perturbative partonic correlation function of the type $<0|{\cal O}^\dagger_B(x'){\cal O}_B(x'')|0>$ in QCD is given by
\bea
&& <0|{\cal O}^\dagger_B(x'){\cal O}_B(x'')|0>=\frac{1}{Z[0]}\int [d{\bar \psi}_U][d\psi_U] [d{\bar \psi}_D] [d\psi_D]\times {\cal O}^\dagger_B(x'){\cal O}_B(x'')\nonumber \\
&& \times {\rm det}[\frac{\delta G_F^d}{\delta \omega^c}] \times {\rm exp}[i\int d^4x [-\frac{1}{4} F_{\mu \lambda}^b(x)F^{\mu \lambda b}(x)-\frac{1}{2\alpha} [G_F^b(x)]^2 \nonumber \\
&& +{\bar \psi}^j_U(x)[\delta^{jl}(i{\not \partial} -m_U)+gT^b_{jl}\aslash^b(x)]\psi^l_U(x) +{\bar \psi}^j_D(x)[\delta^{jl}(i{\not \partial} -m_D)+gT^b_{jl}\aslash^b(x)]\psi^l_D(x)]]\nonumber \\
\label{eb}
\eea
where $Z[0]$ is given by eq. (\ref{ab}), the partonic operator ${\cal O}_B(x)$ is given by eq. (\ref{cb}) and $|0>$ is the vacuum state of the full QCD (not of the pQCD).

The time evolution of the operator ${\cal O}_B(x)$ in the Heisenberg representation is given by
\bea
{\cal O}_B(t,{\vec r}) = e^{-iH~t} {\cal O}_B(t,{\vec r}) e^{iH~t}
\label{fb}
\eea
where $H$ is the QCD hamiltonian of the partons. For the baryon/antibaryon $B$ of momentum ${\vec k}$ the complete set of energy-momentum eigenstates $|P_l({\vec k})>$ is given by
\bea
\sum_{l} |P_l({\vec k})><P_l({\vec k})|=1.
\label{gb}
\eea

The energy-momentum tensor density operator in QCD is given by
\bea
&&T^{\lambda \nu}(x)=
 F^{\lambda \mu b}(x) F_{\mu}^{~\nu b}(x) +\frac{1}{4} g^{ \lambda \nu} F_{\mu \sigma}^b(x)F^{\mu \sigma b}(x)
+ {\bar { \psi}}_U^j(x) \gamma^\lambda [\delta^{jk}i\partial^\nu -igT^b_{jk}A^{\nu b}(x)]{ \psi}_U^k(x)\nonumber \\
&& + {\bar { \psi}}_D^j(x) \gamma^\lambda [\delta^{jk} i\partial^\nu -igT^b_{jk}A^{\nu b}(x)]{ \psi}_D^k(x).
\label{mb}
\eea
From the continuity equation
\bea
\partial_\lambda T^{\lambda \nu}(x) =0
\label{hb}
\eea
we find \cite{nkbr}
\bea
&&\frac{dE({\vec k},t)}{dt} =-\frac{dE_S({\vec k},t)}{dt} \neq 0, ~~~~~~~~~~ E({\vec k},t) =<B({\vec k})|\sum_{q,{\bar q},g}\int d^3r T^{00}(t,{\vec r})|B({\vec k})>, \nonumber \\
&& \frac{dE_S({\vec k},t)}{dt}=<B({\vec k})|\sum_{q,{\bar q},g}\int d^3r \partial_j T^{j0}(t,{\vec r})|B({\vec k})>\neq 0
\label{kb}
\eea
where
\bea
|B({\vec k})> = |P_0({\vec k})>
\label{lb}
\eea
is the energy-momentum eigenstate of baryon/antibaryon $B$ of momentum ${\vec k}$ in its ground state.

Using eqs. (\ref{fb}) and (\ref{gb}) in (\ref{eb}) we find in the Euclidean time
\bea
\sum_{{\vec r}} e^{i{\vec k} \cdot {\vec r}} <0|{\cal O}^\dagger_B(t,{\vec r}) {\cal O}_B(0)|0> =\sum_l |<P_l({\vec k})|{\cal O}_B|0>|^2\times e^{-\int dt E_l({\vec k},t)}
\label{nb}
\eea
where $\int dt$ is the indefinite integration and $E_l({\vec k},t)$ is the energy of all the partons inside the baryon/antibaryon $B$ of momentum ${\vec k}$ in its $l$th energy level state. In the large time limit (by neglecting the higher energy level contributions) we find
\bea
[\sum_{{\vec r}} e^{i{\vec k} \cdot {\vec r}} <0|{\cal O}^\dagger_B(t,{\vec r}) {\cal O}_B(0)|0>]_{t \rightarrow \infty} = |<B({\vec k})|{\cal O}_B|0>|^2\times e^{-\int dt E({\vec k},t)}
\label{ob}
\eea
where $E({\vec k},t)=E_0({\vec k},t)$ is the energy of all the partons inside the baryon/antibaryon $B$ of momentum ${\vec k}$ in its ground state.

Similar to the derivation of eq. (\ref{ob}) we find from eq. (\ref{kb})
\bea
\frac{dE_S({\vec k},t)}{dt}=\left[\frac{\sum_{{\vec r}'} e^{i{\vec k} \cdot {\vec r}'}<0| {\cal O}^\dagger_B(t',{\vec r}')\sum_{q,{\bar q},g} \int d^3r \partial_j T^{j0}(t,{\vec r}){\cal O}_B(0)|0>} {\sum_{{\vec r}'} e^{i{\vec k} \cdot {\vec r}'}<0| {\cal O}^\dagger_B(t',{\vec r}'){\cal O}_B(0)|0>} \right]_{t' \rightarrow \infty}.
\label{pb}
\eea
The energy $E_B({\vec k})$ of the baryon/antibaryon $B$ of momentum ${\vec k}$ is given by \cite{nkbr}
\bea
E_B({\vec k})=E({\vec k},t)+E_S({\vec k},t)
\label{qb}
\eea
where $E({\vec k},t)$ is the energy of all the partons inside the baryon/antibaryon $B$ as given by eq. (\ref{kb}) and $E_S({\vec k},t)$ is the non-zero boundary term due to confinement in QCD as given by eqs. (\ref{kb}) and (\ref{pb}).

Using eqs. (\ref{qb}) and (\ref{pb}) in (\ref{ob}) we find
\bea
|<B({\vec k})|{\cal O}_B|0>|^2\times e^{-t E_B({\vec k})}=\left[\frac{\sum_{{\vec r}} e^{i{\vec k} \cdot {\vec r}} <0|{\cal O}^\dagger_B(t,{\vec r}) {\cal O}_B(0)|0>}{e^{\int dt \left[\frac{\sum_{{\vec r}'} e^{i{\vec k} \cdot {\vec r}'}<0| {\cal O}^\dagger_B(t',{\vec r}')\sum_{q,{\bar q},g} \int dt \int d^3r \partial_j T^{j0}(t,{\vec r}){\cal O}_B(0)|0>} {\sum_{{\vec r}'} e^{i{\vec k} \cdot {\vec r}'}<0| {\cal O}^\dagger_B(t',{\vec r}'){\cal O}_B(0)|0>} \right]_{t' \rightarrow \infty}}}  \right]_{t \rightarrow \infty}
\label{rb}
\eea
which is the non-perturbative formula of the formation of the baryon/antibaryon $B~(=p,n,{\bar p},{\bar n})$ from the quarks and gluons which can be calculated by using the lattice QCD method where the partonic operator ${\cal O}_B(x)$ is given by eq. (\ref{cb}). It is straightforward to extend the eq. (\ref{rb}) to other baryons/antibaryons.

\section{Baryon and Antibaryon  formation from finite temperature quarks and gluons using imaginary time path integral formalism in equilibrium QCD }

For the quarks and gluons at the finite temperature $T$ the generating functional of QCD at the finite temperature $T$ in the absence of any external sources is given by
\bea
&& Z[0]=\int [d{\bar \psi}_U][d\psi_U] [d{\bar \psi}_D] [d\psi_D] \times {\rm det}[\frac{\delta G_F^d}{\delta \omega^c}] \times {\rm exp}[-\int_0^{\frac{1}{T}} d\tau \int d^3r [-\frac{1}{4} F_{\mu \lambda}^b(\tau,r)F^{\mu \lambda b}(\tau,r)\nonumber \\
&&-\frac{1}{2\alpha} [G_F^b(\tau,r)]^2 +{\bar \psi}^j_U(\tau,r)[\delta^{jl}(i{\not \partial} -m_U)+gT^b_{jl}\aslash^b(\tau,r)]\psi^l_U(\tau,r) +{\bar \psi}^j_D(\tau,r)[\delta^{jl}(i{\not \partial} -m_D)\nonumber \\
&&+gT^b_{jl}\aslash^b(\tau,r)]\psi^l_D(\tau,r)]]
\label{abm}
\eea
where $\tau$ is the Euclidean time and the quark, gluon fields satisfy the periodic boundary conditions
\bea
\psi^j_U(\tau,r)=\psi^j_U(\tau+\frac{1}{T},r),~~~~~~~~\psi^j_D(\tau,r)=\psi^j_D(\tau+\frac{1}{T},r),~~~~~~~~~A^b_\lambda(\tau,r)=A^b_\lambda(\tau+\frac{1}{T},r).\nonumber \\
\eea

The expectation value of the two point non-perturbative partonic correlation function of the type $<in|{\cal O}^\dagger_B(\tau'',r''){\cal O}_B(0)|in>$ in QCD at the finite temperature $T$ is given by
\bea
&& <in|{\cal O}^\dagger_B(\tau'',r''){\cal O}_B(0)|in>=\frac{1}{Z[0]}\int [d{\bar \psi}_U][d\psi_U] [d{\bar \psi}_D] [d\psi_D]\times {\cal O}^\dagger_B(\tau'',r''){\cal O}_B(0)\nonumber \\
&& \times {\rm det}[\frac{\delta G_F^d}{\delta \omega^c}] \times {\rm exp}[-\int_0^{\frac{1}{T}} d\tau \int d^3r [-\frac{1}{4} F_{\mu \lambda}^b(\tau,r)F^{\mu \lambda b}(\tau,r)-\frac{1}{2\alpha} [G_F^b(\tau,r)]^2 \nonumber \\
&& +{\bar \psi}^j_U(\tau,r)[\delta^{jl}(i{\not \partial} -m_U)+gT^b_{jl}\aslash^b(\tau,r)]\psi^l_U(\tau,r) +{\bar \psi}^j_D(\tau,r)[\delta^{jl}(i{\not \partial} -m_D)+gT^b_{jl}\aslash^b(\tau,r)]\psi^l_D(\tau,r)]]\nonumber \\
\label{ebm}
\eea
where $Z[0]$ is given by eq. (\ref{abm}), the $|in>$ is the ground state of the full QCD  at the finite temperature $T$ (not of the pQCD at the finite temperature $T$) and the partonic operator ${\cal O}_B(x)$ is given by eq. (\ref{cb}) where $B=p,n,{\bar p},{\bar n}$=proton, neutron, antiproton, antineutron.

Note that unlike
\bea
<0|e^{-iHt}=0
\label{vo}
\eea
in eq. (\ref{nb}) in QCD in vacuum we find in QCD at the finite temperature that
\bea
<in|e^{H\tau}\neq 0.
\label{fbm}
\eea
Hence using eqs. (\ref{fb}), (\ref{fbm}) and (\ref{gb}) in (\ref{ebm}) we find
\bea
\sum_{{\vec r}''} e^{i{\vec k} \cdot {\vec r}''} <in|e^{-H\tau''}{\cal O}^\dagger_B(\tau'',{\vec r}'') {\cal O}_B(0)|in> =\sum_l |<P_l({\vec k})|{\cal O}_B|in>|^2\times e^{-\int d\tau'' E_l({\vec k},\tau'')}\nonumber \\
\label{nbm}
\eea
where $\int d\tau''$ is the indefinite integration and $E_l({\vec k},\tau'')$ is the energy of all the partons inside the baryon/antibaryon $B$ of momentum ${\vec k}$ in its $l$th energy level state.

Note that the baryon/antibaryon $B$ (such as proton, neutron, antiproton, antineutron) belongs to the confined phase of QCD whereas the QCD at the finite temperature (the quark-gluon plasma) belongs to the de-confined phase of QCD. Hence the baryon/antibaryon $B$ is not formed inside the quark-gluon plasma medium but the baryon/antibaryon $B$ is formed outside the quark-gluon plasma medium, {\it i. e.}, the baryon/antibaryon $B$ is formed in the vacuum. Hence $\tau'' \rightarrow \infty$ can be taken in eq. (\ref{nbm}) even if the upper limit of $\tau$ in eq. (\ref{ebm}) is $\frac{1}{T}$.

Hence in the large time limit (by neglecting the higher energy level contributions) we find
\bea
[\sum_{{\vec r}''} e^{i{\vec k} \cdot {\vec r}''} <in|e^{-H\tau''}{\cal O}^\dagger_B(\tau'',{\vec r}'') {\cal O}_B(0)|in>]_{\tau'' \rightarrow \infty} = |<B({\vec k})|{\cal O}_B|in>|^2\times e^{-\int d\tau'' E({\vec k},\tau'')}\nonumber \\
\label{obm}
\eea
where $E({\vec k},\tau'')=E_0({\vec k},\tau'')$ is the energy of all the partons inside the baryon/antibaryon $B$ of momentum ${\vec k}$ in its ground state.

Since the baryon/antibaryon $B$ is formed in vacuum we can use eqs. (\ref{qb}) and (\ref{pb}) in (\ref{obm}) to find
\bea
&& |<B({\vec k})|{\cal O}_B|in>|^2=\left[\frac{\sum_{{\vec r}''} e^{i{\vec k} \cdot {\vec r}''} <in|e^{-H\tau''}{\cal O}^\dagger_B(\tau'',{\vec r}'') {\cal O}_B(0)|in>\times e^{\tau'' E_B({\vec k})}}{e^{\int d\tau'' \left[\frac{\sum_{{\vec r}'} e^{i{\vec k} \cdot {\vec r}'}<0| {\cal O}^\dagger_B(\tau',{\vec r}')\sum_{q,{\bar q},g}\int d\tau''  \int d^3r'' \partial_j T^{j0}(\tau'',{\vec r}''){\cal O}_B(0)|0>} {\sum_{{\vec r}'} e^{i{\vec k} \cdot {\vec r}'}<0| {\cal O}^\dagger_B(\tau',{\vec r}'){\cal O}_B(0)|0>} \right]_{\tau' \rightarrow \infty}}}  \right]_{\tau'' \rightarrow \infty}
\label{rbm}
\eea
which is the non-perturbative formula of the baryon/antibaryon $B~(=p,n,{\bar p},{\bar n})$ formation from the finite temperature quarks and gluons which can be calculated by using lattice QCD method  at the finite temperature $T$ where the partonic operator ${\cal O}_B(x)$ is given by eq. (\ref{cb}). It is straightforward to extend the eq. (\ref{rbm}) to other baryons/antibaryons.

\section{Baryon and Antibaryon formation from Non-Equilibrium quarks and gluons in the early universe Using Closed-Time Path Integral Formalism in non-equilibrium QCD }

In the previous section we have studied the baryon/antibaryon formation from the finite temperature quarks and gluons in QCD in equilibrium by implementing the non-vanishing boundary term due to confinement in QCD. However, as mentioned in the introduction, the baryon asymmetry of the universe cannot be explained if the early universe was in equilibrium just after the big bang \cite{bs2}. Unlike QCD at the finite temperature $T$, one can not define the temperature $T$ in non-equilibrium QCD.

In this section we study the baryons (proton and neutron) and the antibaryons (antiproton and antineutron) formation from the non-equilibrium quarks and gluons in QCD in the early universe just after the big bang from the first principle in QCD by using the closed-time path integral formalism in non-equilibrium QCD. The extension of this formalism to other baryons/antibaryons is straightforward.

For the quarks and gluons in non-equilibrium the generating functional of non-equilibrium QCD [by using the closed-time path integral formalism] in the absence of any external sources is given by \cite{ne,ne1,ne2,ne3}
\bea
&& Z[0]=\int [d{\bar \psi}_{U+}][d\psi_{U+}] [d{\bar \psi}_{D+}] [d\psi_{D+}][d{\bar \psi}_{U-}][d\psi_{U-}] [d{\bar \psi}_{D-}] [d\psi_{D-}] \times {\rm det}[\frac{\delta G_{F+}^d}{\delta \omega^c_+}] \times {\rm det}[\frac{\delta G_{F-}^d}{\delta \omega^c_-}] \nonumber \\
&& \times {\rm exp}[i\int d^4x [-\frac{1}{4} F_{\mu \lambda +}^b(x)F^{\mu \lambda b}_+(x)+\frac{1}{4} F_{\mu \lambda -}^b(x)F^{\mu \lambda b}_-(x)-\frac{1}{2\alpha} [G_{F+}^b(x)]^2 +\frac{1}{2\alpha} [G_{F-}^b(x)]^2\nonumber \\
&& +{\bar \psi}^j_{U+}(x)[\delta^{jl}(i{\not \partial} -m_U)+gT^b_{jl}\aslash^b_+(x)]\psi^l_{U+}(x) +{\bar \psi}^j_{D+}(x)[\delta^{jl}(i{\not \partial} -m_D)+gT^b_{jl}\aslash^b_+(x)]\psi^l_{D+}(x)\nonumber \\
&& -{\bar \psi}^j_{U-}(x)[\delta^{jl}(i{\not \partial} -m_U)+gT^b_{jl}\aslash^b_-(x)]\psi^l_{U-}(x) -{\bar \psi}^j_{D-}(x)[\delta^{jl}(i{\not \partial} -m_D)+gT^b_{jl}\aslash^b_-(x)]\psi^l_{D-}(x)]]\nonumber \\
&& \times <A_+,{\bar \psi}_{U+},\psi_{U+},{\bar \psi}_{D+},\psi_{D+},0|\rho|0,\psi_{D-},{\bar \psi}_{D-},\psi_{U-},{\bar \psi}_{U-},A_->
\label{abn}
\eea
where $\pm$ is the closed-time path index and $\rho$ is the initial density of states of the partons in non-equilibrium QCD.

The expectation value of the two point non-perturbative partonic correlation function of the type $<in|{\cal O}^{\dagger B}_+(x'){\cal O}^B_+(x'')|in>$ in non-equilibrium QCD is given by
\bea
&& <in|{\cal O}^{\dagger B}_+(x'){\cal O}^B_+(x'')|in>=\int [d{\bar \psi}_{U+}][d\psi_{U+}] [d{\bar \psi}_{D+}] [d\psi_{D+}][d{\bar \psi}_{U-}][d\psi_{U-}] [d{\bar \psi}_{D-}] [d\psi_{D-}] \nonumber \\
&&\times {\cal O}^{\dagger B}_+(x'){\cal O}^B_+(x'') \times {\rm det}[\frac{\delta G_{F+}^d}{\delta \omega^c_+}] \times {\rm det}[\frac{\delta G_{F-}^d}{\delta \omega^c_-}] \nonumber \\
&& \times {\rm exp}[i\int d^4x [-\frac{1}{4} F_{\mu \lambda +}^b(x)F^{\mu \lambda b}_+(x)+\frac{1}{4} F_{\mu \lambda -}^b(x)F^{\mu \lambda b}_-(x)-\frac{1}{2\alpha} [G_{F+}^b(x)]^2 +\frac{1}{2\alpha} [G_{F-}^b(x)]^2\nonumber \\
&& +{\bar \psi}^j_{U+}(x)[\delta^{jl}(i{\not \partial} -m_U)+gT^b_{jl}\aslash^b_+(x)]\psi^l_{U+}(x) +{\bar \psi}^j_{D+}(x)[\delta^{jl}(i{\not \partial} -m_D)+gT^b_{jl}\aslash^b_+(x)]\psi^l_{D+}(x)\nonumber \\
&& -{\bar \psi}^j_{U-}(x)[\delta^{jl}(i{\not \partial} -m_U)+gT^b_{jl}\aslash^b_-(x)]\psi^l_{U-}(x)- {\bar \psi}^j_{D-}(x)[\delta^{jl}(i{\not \partial} -m_D)+gT^b_{jl}\aslash^b_-(x)]\psi^l_{D-}(x)]]\nonumber \\
&& \times <A_+,{\bar \psi}_{U+},\psi_{U+},{\bar \psi}_{D+},\psi_{D+},0|\rho|0,\psi_{D-},{\bar \psi}_{D-},\psi_{U-},{\bar \psi}_{U-},A_->
\label{ebn}
\eea
where $|in>$ is the ground state of the full QCD in non-equilibrium (not of the pQCD in non-equilibrium) and the partonic operator ${\cal O}^B_+(x)$ [where
$B=p,n,{\bar p},{\bar n}$= proton, neutron, antiproton, antineutron] is given by
\bea
&& {\cal O}^p_+(x) =\epsilon^{jlk} \psi^{jT}_{U+}C\gamma_5 \psi^l_{D+} \psi^k_{U+}(x),~~~~~~~~~~~~~~~~{\cal O}^n_+(x) =\epsilon^{jlk} \psi^{jT}_{D+}C\gamma_5 \psi^l_{U+} \psi^k_{D+}(x), \nonumber \\
&& {\cal O}^{\bar p}_+(x) =\epsilon^{jlk} {{\bar \psi}^{jT}_{U+}}C\gamma_5 {\bar \psi}^l_{D+} {\bar \psi}^k_{U+}(x),~~~~~~~~~~~~~~~~{\cal O}^{\bar n}_+(x) =\epsilon^{jlk} {{\bar \psi}^{jT}_{D+}}\gamma_5 {\bar \psi}^l_{U+} {\bar \psi}^k_{D+}(x).
\label{cbn}
\eea
Note that unlike
\bea
<0|e^{-iHt}=0
\label{von}
\eea
in eq. (\ref{nb}) in QCD in vacuum we find in non-equilibrium QCD that
\bea
<in|e^{-iHt}\neq 0.
\label{fbn}
\eea
As mentioned earlier, unlike QCD at the finite temperature $T$, one can not define a temperature $T$ in non-equilibrium QCD.
In addition to this the QCD at the finite temperature $T$ in section III is formulated in the imaginary time formalism (the Euclidean time formalism) where the Euclidean time can be related to the inverse of the temperature $T$. However, the imaginary time formalism (the Euclidean time formalism) does not work in non-equilibrium QCD because the non-equilibrium QCD is formulated in the real time (in the Minkowski time) by using in the closed-time path integral formalism \cite{ne,ne1,ne2,ne3}.

Hence using eqs. (\ref{fb}) and (\ref{gb}) in (\ref{ebn}) we find
\bea
\sum_{{\vec r}} e^{i{\vec k} \cdot {\vec r}} <in|{\cal O}^{\dagger B}_+(0,{\vec r}) e^{-tH} {\cal O}^B_+(0)|in> =\sum_l |<P_l({\vec k})|{\cal O}^B_+|in>|^2\times e^{-\int dt E_l({\vec k},t)}
\label{nbn}
\eea
where $\int dt$ is the indefinite integration and $E_l({\vec k},t)$ is the energy of all the partons inside the baryon/antibaryon $B$ of momentum ${\vec k}$ in its $l$th energy level state. In the large time limit (by neglecting the higher energy level contributions) we find
\bea
[\sum_{{\vec r}} e^{i{\vec k} \cdot {\vec r}} <in|{\cal O}^{\dagger B}_+(0,{\vec r}) e^{-tH}{\cal O}^B_+(0)|in>]_{t \rightarrow \infty} = |<B({\vec k})|{\cal O}^B_+|in>|^2\times e^{-\int dt E({\vec k},t)}
\label{obn}
\eea
where $E({\vec k},t)=E_0({\vec k},t)$ is the energy of all the partons inside the baryon/antibaryon $B$ of momentum ${\vec k}$ in its ground state.

Note that the baryon/antibaryon $B$ (such as proton, neutron, antiproton, antineutron) belongs to the confined phase of QCD whereas the non-equilibrium QCD medium (the non-equilibrium quark-gluon plasma) belongs to the de-confined phase of QCD. Hence the baryon/antibaryon $B$ is not formed inside the non-equilibrium quark-gluon plasma medium but the baryon/antibaryon $B$ is formed outside the non-equilibrium quark-gluon plasma medium, {\it i. e.}, the baryon/antibaryon $B$ is formed in the vacuum.

Since the baryon/antibaryon $B$ is formed in the vacuum we can use eqs. (\ref{qb}) and (\ref{pb}) in (\ref{obn}) to find
\bea
|<B({\vec k})|{\cal O}^B_+|in>|^2=\left[\frac{\sum_{{\vec r}} e^{i{\vec k} \cdot {\vec r}} <in|{\cal O}^{\dagger B}_+(0,{\vec r})e^{-tH} {\cal O}^B_+(0)|in>\times e^{t E_B({\vec k})}}{e^{\int dt \left[\frac{\sum_{{\vec r}'} e^{i{\vec k} \cdot {\vec r}'}<0| {\cal O}^\dagger_B(t',{\vec r}')\sum_{q,{\bar q},g} \int dt \int d^3r \partial_j T^{j0}(t,{\vec r}){\cal O}_B(0)|0>} {\sum_{{\vec r}'} e^{i{\vec k} \cdot {\vec r}'}<0| {\cal O}^\dagger_B(t',{\vec r}'){\cal O}_B(0)|0>} \right]_{t' \rightarrow \infty}}}  \right]_{t \rightarrow \infty}
\label{rbn}
\eea
which is the non-perturbative formula of the baryon/antibaryon $B~(=p,n,{\bar p},{\bar n}$) formation from the non-equilibrium quarks and gluons derived from the first principle in QCD by using the closed-time path integral formalism in non-equilibrium QCD where the partonic operator ${\cal O}^B_+(x)$ is given by eq. (\ref{cbn}). It is straightforward to extend the eq. (\ref{rbn}) to other baryons/antibaryons.

\section{Conclusions}
Baryon number violation, CP violation and non-equilibrium evolution of the early universe just after the big bang are proposed in the literature to be necessary conditions to explain the observed matter-antimatter asymmetry (baryon asymmetry) of the universe. Since the free quarks are not observed and the baryons are formed by the confinement of quarks and gluons, it is necessary to understand how the baryons were formed for the first time from the non-equilibrium quarks and gluons in the early universe just after the big bang in order to study the baryon asymmetry of the universe. In this paper we have studied the baryon formation from the non-equilibrium quarks and gluons in the early universe just after the big bang from the first principle in QCD by using the closed-time path integral formalism in non-equilibrium QCD.

The baryon and antibaryon production (such as proton and antiproton production) are experimentally measured in the Pb-Pb collisions by the ALICE collaborations at the high energy heavy-ion colliders at LHC \cite{lh1,lh2}. Hence the baryons/antibaryons production from the  non-equilibrium quark-gluon plasma at RHIC and LHC \cite{qgf,qgf1,qgf2} experiments in the laboratory can provide useful information about the baryons/antibaryons formation from the non-equilibrium quark-gluon plasma in the early universe just after $\sim 10^{-4}$ seconds of the big bang.

\end{document}